# Hydrodynamic construction of the electromagnetic field

**Peter Holland**

Green College
University of Oxford
Oxford OX2 6HG
England

peter.holland@green.ox.ac.uk

11th June 2005

## Abstract

We present an alternative Eulerian hydrodynamic model for the electromagnetic field in which the discrete vector indices in Maxwell's equations are replaced by continuous angular freedoms, and develop the corresponding Lagrangian picture in which the fluid particles have rotational and translational freedoms. This enables us to extend to the electromagnetic field the exact method of state construction proposed previously for spin 0 systems, in which the time-dependent wavefunction is computed from a single-valued continuum of deterministic trajectories where two spacetime points are linked by at most a single orbit. The deduction of Maxwell's equations from continuum mechanics is achieved by generalizing the spin 0 theory to a general Riemannian manifold from which the electromagnetic construction is extracted as a special case. In particular, the flat-space Maxwell equations are represented as a curved-space Schrödinger equation for a massive system. The Lorentz covariance of the Eulerian field theory is obtained from the non-covariant Lagrangian-coordinate model as a kind of collective effect. The method makes manifest the electromagnetic analogue of the quantum potential that is tacit in Maxwell's equations. This implies a novel definition of the "classical limit" of Maxwell's equations that differs from geometrical optics. It is shown that Maxwell's equations may be obtained by canonical quantization of the classical model. Using the classical trajectories a novel expression is derived for the propagator of the electromagnetic field in the Eulerian picture. The trajectory and propagator methods of solution are illustrated for the case of a light wave.

 "…it is a good thing to have two ways of looking at a subject, and to admit that there *are* two ways of looking at it." [1]

## 1. INTRODUCTION

In a recent article [2] a method was described that provides an exact scheme to calculate the time-dependent wavefunction for a spin 0 system from a single-valued continuum of deterministic trajectories where two spacetime points are linked by at most a single orbit. A natural language for the theory is offered by the hydrodynamic analogy, in which wave mechanics corresponds to the Eulerian picture and the trajectory theory to the Lagrangian picture. The method entails the derivation of the time-dependent Schrödinger equation from a continuum mechanics model in which the fluid elements are subject to a specific interaction potential, and shows that the deterministic trajectory concept may be regarded as a basic component of the quantum description (and potentially of other wave theories) and not just of one of its interpretations (e.g., the de Broglie-Bohm model). Our goal here is to extend this method of deduction of the wave equation to include spin. Specifically, we show how the relativistic spin 1 field obeying the source-free Maxwell equations can be computed from the Lagrangian trajectories in an extended hydrodynamic model in which the fluid particles acquire internal rotational freedoms in addition to translational ones.

An Eulerian hydrodynamic picture for Maxwell's equations has been developed previously [3,4], employing a method similar to that used for spin $\frac{1}{2}$ systems (e.g., [5] and references in [6]). However, the equations in this "standard" Eulerian approach to systems with spin are rather complicated and difficult to interpret, at least in comparison with the simplicity of the spin 0 Madelung model, and there are ambiguities in identifying suitably defined phases of the relevant wavefunctions (and the associated vortex structures) [7]. In addition, in connection with our programme, it is not clear how to develop a suitable Lagrangian-coordinate version of the theory. The basic conservation law in the standard hydrodynamic model of Maxwell's equations is Poynting's theorem and a natural definition of the flow lines would be the integral curves of the Poynting vector divided by the energy density. Undoubtedly, these paths provide insight into electromagnetic phenomena [6] (e.g., two-slit interference [8]) but it is clear that an ensemble of such paths would not contain sufficient information to construct the six field components according to the method set out in [2].

It has been suggested that the origin of the problem of the complexity of the spin $\frac{1}{2}$ hydrodynamic equations is that the standard approach works with a (angular momentum) representation of the quantum theory in which the rotational freedoms appear as discrete indices in the wavefunction [6]. This observation applies equally to spin 1. The local fluid quantities (density, velocity, spin vector,...) are defined by "averaging" over these indices, and reproducing through them all the information in the wavefunction requires introducing ever more complex quantities and combinations of quantities (especially in the many-particle case for matter theories; see, e.g., [9]). The alternative procedure advocated in [6] for spin $\frac{1}{2}$ and used here is to start from the angular coordinate representation in which the spin freedoms are represented as continuous parameters $\alpha$ (Euler angles) in the wavefunction, on the same footing as the spatial variables $x$: $\psi(x,\alpha,t)$. This implies a physically clearer and simpler hydrodynamic-like model, in both its Eulerian and Lagrangian guises. The phase $S$ of the wavefunction is immediately identifiable and the equations for the fluid paths are defined in terms of the gradients of $S$ with respect to the coordinates, obvious generalizations of the spin 0 theory. The approach also provides a natural framework to study vortices, and it has the merit of locating spin analogues of the spin 0 quantum potential, especially in the electromagnetic application. Regarding Maxwell's theory as the first quantized theory of light we would expect something akin to the quantum potential to appear in the field equations, but in the usual representation this is hidden.

The final advantage of the angular coordinate approach is that it allows us to extend our method of wavefunction construction to the electromagnetic field. The deduction of Maxwell's equations from continuum mechanics is achieved by first generalizing the theory of [2] to a general Riemannian manifold from which the electromagnetic construction is extracted as a special case. In particular, the flat space Maxwell equations are represented as a curved-space Schrödinger equation for a massive system. The Lagrangian-coordinate model presented here is not relativistically covariant and one of our results is that the Lorentz covariance of a (Eulerian) field theory can be derived from a non-covariant theory as a kind of collective effect. The limiting case of negligible quantum potential yields a "classical limit" of electromagnetic theory that differs from geometrical optics. It is shown that Maxwell's equations may be obtained by canonical quantization of the classical model. Using the classical trajectories a novel expression may be derived for the propagator of the electromagnetic field in the Eulerian picture.

## 2. HYDRODYNAMIC FORMULATION OF MAXWELL'S EQUATIONS

### A. Maxwell's equations in Schrödinger form

The source-free Maxwell equations in free space are

$$\varepsilon_{ijk}\partial_j E_k = -\frac{\partial B_i}{\partial t}, \quad \varepsilon_{ijk}\partial_j B_k = \frac{1}{c^2}\frac{\partial E_i}{\partial t} \tag{2.1}$$

$$\partial_i E_i = 0, \quad \partial_i B_i = 0 \tag{2.2}$$

where $\partial_i = \partial/\partial x_i$, $x_i = (x,y,z)$, $\varepsilon_{ijk}$ is the completely antisymmetric symbol with $\varepsilon_{123} = 1$, and $i,j,k,\ldots = 1,2,3$. Summation over repeated indices is always assumed. The time derivatives of relations (2.2) being derivable from (2.1), we shall regard (2.2) as constraints on the fields at $t = 0$ rather than as dynamical conditions.

Our method proceeds by first expressing Maxwell's equations (2.1) in Schrödinger form. There are several ways of achieving this (e.g., [3, 4, 10-17]). In the version used here we start by representing the field in terms of the Riemann-Silberstein 3-vector $F_i = \sqrt{\varepsilon_0/2}\,(E_i + icB_i)$ [3, 4, 12] and the rotational aspects of the field equations (i.e., the curls) in terms of the 3x3 angular momentum matrices $(s_i)_{jk} = -i\hbar\varepsilon_{ijk}$, which obey the commutation relations

$$\left[s_i, s_j\right] = i\hbar\varepsilon_{ijk}s_k. \tag{2.3}$$

Eqs. (2.1) are then equivalent to the equations

$$i\hbar\frac{\partial F_i}{\partial t} = -ic\left(s_j\right)_{ik}\partial_j F_k \tag{2.4}$$

subject to $\partial_i F_i = 0$ at $t = 0$ (which we shall henceforth always assume is obeyed). This is the required Schrödinger form. The role of $\hbar$ is simply to provide a constant having the dimension of action that allows the hydrodynamic decomposition of Maxwell's equations – naturally, it plays no role in (2.1), and drops out of the computations of the trajectories.

To lay the groundwork for the passage to the continuous representation of the spin freedoms, we transform to a representation of the angular momentum matrices in which the $z$-component is diagonal. This is effected by the unitary matrix

$$U_{ai} = \frac{1}{\sqrt{2}}\begin{pmatrix} -1 & i & 0 \\ 0 & 0 & \sqrt{2} \\ 1 & i & 0 \end{pmatrix}. \tag{2.5}$$

Maxwell's equations are now

$$i\hbar\frac{\partial G_a}{\partial t} = -ic\big(J_j\big)_{ab}\partial_j G_b \tag{2.6}$$

where $G_a = U_{ai}F_i$, $J_i = Us_iU^{-1}$ and $a,b = +1,0,-1$. We have

$$\begin{pmatrix} G_1 \\ G_0 \\ G_{-1} \end{pmatrix} = \frac{1}{\sqrt{2}}\begin{pmatrix} -F_1 + iF_2 \\ \sqrt{2}F_3 \\ F_1 + iF_2 \end{pmatrix} \quad \text{or} \quad \begin{pmatrix} F_1 \\ F_2 \\ F_3 \end{pmatrix} = \frac{1}{\sqrt{2}}\begin{pmatrix} G_{-1} - G_1 \\ -iG_{-1} - iG \\ \sqrt{2}G_0 \end{pmatrix} \tag{2.7}$$

and

$$\big(J_1\big)_{ab} = \frac{\hbar}{\sqrt{2}}\begin{pmatrix} 0 & 1 & 0 \\ 1 & 0 & 1 \\ 0 & 1 & 0 \end{pmatrix}, \quad \big(J_2\big)_{ab} = \frac{\hbar}{\sqrt{2}}\begin{pmatrix} 0 & -i & 0 \\ i & 0 & -i \\ 0 & i & 0 \end{pmatrix}, \quad \big(J_3\big)_{ab} = \hbar\begin{pmatrix} 1 & 0 & 0 \\ 0 & 0 & 0 \\ 0 & 0 & -1 \end{pmatrix}. \tag{2.8}$$

## B. Continuous representation

The final step consists in passing to the angular coordinate representation. Using the definition of the Euler angles $(\alpha_r) = (\alpha,\beta,\gamma)$, $r = 1,2,3$, and conventions of [6], the angular momentum components become differential operators:

$$\left.\begin{array}{l} \hat{M}_1 = i\hbar\big(\cos\beta\partial_\alpha - \sin\beta\cot\alpha\partial_\beta + \sin\beta\mathrm{cosec}\alpha\partial_\gamma\big) \\ \hat{M}_2 = i\hbar\big(\text{-}\sin\beta\partial_\alpha - \cos\beta\cot\alpha\partial_\beta + \cos\beta\mathrm{cosec}\alpha\partial_\gamma\big) \\ \hat{M}_3 = i\hbar\partial_\beta \end{array}\right\} \tag{2.9}$$

where $\partial_\alpha = \partial/\partial\alpha$ etc. The Schrödinger equation (2.6) becomes

$$i\hbar\frac{\partial\psi(x,\alpha)}{\partial t} = -ic\hat{M}_i\partial_i\psi(x,\alpha) \tag{2.10}$$

or, introducing the real operator $\hat{\lambda}_i = \hat{M}_i/(-i\hbar)$,

$$i\hbar\frac{\partial\psi}{\partial t} = -c\hbar\hat{\lambda}_i\partial_i\psi \tag{2.11}$$

where $\psi$ is a function on the six-dimensional manifold $M = \Re^3 \otimes SO(3)$ whose points are labelled by $(x,\alpha)$. In this representation the wavefunction may be expanded in terms of an orthonormal set of spin 1 basis functions $u_a(\alpha)$ (eigenfunctions of the anomalous angular momentum operator $\hat{M}'_3$ corresponding to eigenvalue zero – see [6]):

$$\psi(x,\alpha,t) = G_a(x,t)u_a(\alpha), \quad a = 1,0,-1, \tag{2.12}$$

where

$$u_1(\alpha) = \big(\sqrt{3}/4\pi\big)\sin\alpha e^{-i\beta}, \quad u_0(\alpha) = i\big(\sqrt{3}/2\sqrt{2}\pi\big)\cos\alpha, \quad u_{-1}(\alpha) = \big(\sqrt{3}/4\pi\big)\sin\alpha e^{i\beta}, \tag{2.13}$$

with

$$\int u_a^{\ *}(\alpha)u_b(\alpha)d\Omega = \delta_{ab}, \quad d\Omega = \sin\alpha d\alpha d\beta d\gamma, \quad \alpha \in [0,\pi], \beta \in [0,2\pi], \gamma \in [0,2\pi]. \quad (2.14)$$

It is readily checked that

$$\int u_a^{\ *}(\alpha)\hat{M}_i u_b(\alpha)d\Omega = (J_i)_{ab} \quad (2.15)$$

and multiplying (2.10) by $u_a^{\ *}(\alpha)$ and using (2.14) we recover Maxwell's equations in the form (2.6).

In this formalism the field equations (2.10) come out as second-order partial differential equations, and summation over $i$ or $a$ is replaced by integration over $\alpha_r$. For example, for the energy density and the Poynting vector we have the alternate expressions

$$\frac{\varepsilon_0}{2}\left(\mathbf{E}^2 + c^2\mathbf{B}^2\right) = F_i^{\ *}F_i = G_a^{\ *}G_a = \int |\psi(x,\alpha)|^2 d\Omega \quad (2.16)$$

$$\varepsilon_0 c^2(\mathbf{E}\times\mathbf{B})_i = (c/\hbar)F_j^{\ *}(s_i)_{jk}F_k = (c/\hbar)G_a^{\ *}(J_i)_{ab}G_b = \frac{c}{\hbar}\int \psi^*(x,\alpha)\hat{M}_i\psi(x,\alpha)d\Omega. \quad (2.17)$$

To obtain the hydrodynamic model we follow Madelung [18] and express the wavefunction in polar form: $\psi = \sqrt{\rho}\exp(iS/\hbar)$, where $\rho$ has the dimension of energy density in $M$. Splitting the wave equation (2.11) into real and imaginary parts then gives the relations

$$\frac{\partial S}{\partial t} + \frac{c}{\hbar}\hat{\lambda}_i S\partial_i S + Q = 0 \quad (2.18)$$

$$\frac{\partial \rho}{\partial t} + \frac{c}{\hbar}\partial_i\left(\rho\hat{\lambda}_i S\right) + \frac{c}{\hbar}\hat{\lambda}_r(\rho\partial_i S) = 0 \quad (2.19)$$

where

$$Q = -c\hbar\frac{\hat{\lambda}_i\partial_i\sqrt{\rho}}{\sqrt{\rho}}. \quad (2.20)$$

These two equations are equivalent to Maxwell's equations (2.1), subject to the proviso that $\rho$ and $S$ obey certain conditions inherited from $\psi$. In particular, the single-valuedness of the wavefunction requires

$$\oint_{C_0} \partial_i S dx_i + \partial_r S d\alpha_r = nh, \quad n \in Z, \quad (2.21)$$

where $C_0$ is a closed curve in $M$. In the hydrodynamic model the number $|n|$ is interpreted as the net *strength* of the vortices contained in $C$. These occur in nodal regions ($\psi = 0$) where $S$ is singular.

Comparing (2.19) with the Eulerian continuity equation corresponding to a fluid of density $\rho$ with translational and rotational freedoms,

$$\frac{\partial \rho}{\partial t} + \partial_i(\rho v_i) + \hat{\lambda}_i(\rho\omega_i) = 0, \quad (2.22)$$

we shall make the following identifications for the velocity and angular velocity fields[1]:

$$v_i = (c/\hbar)\hat{\lambda}_i S, \quad \omega_i = (c/\hbar)\partial_i S. \tag{2.23}$$

Clearly, we obtain a kind of potential flow (strictly, quasi-potential in view of (2.21)), the potential being $(c/\hbar)S$. Note, however, the unorthodox connections between the potential and the hydrodynamic quantities – $v_i$ ($\omega_i$) is a gradient with respect to the angular (spatial) variables. The quantity $Q$ in (2.18) is the analogue for Maxwell's equations of the quantum potential that appears in the polar decomposition of the Schrödinger equation for a massive particle. As we shall see in Sec. 4, $Q$ has the classic form "$-\nabla^2\sqrt{\rho}/\sqrt{\rho}$" when the appropriate metric on $M$ is identified.

From the Bernoulli-like (or Hamilton-Jacobi-like) equation (2.18) we may obtain the analogue(s) of Euler's force law for the electromagnetic fluid. Applying first $\partial_i$, rearranging and using (2.23) we get

$$\left(\frac{\partial}{\partial t} + v_j\partial_j + \omega_j\hat{\lambda}_j\right)\omega_i = -\frac{c}{\hbar}\partial_i Q. \tag{2.24}$$

Next, applying $\hat{\lambda}_i$ and using $\left[\hat{\lambda}_i,\hat{\lambda}_j\right] = -\varepsilon_{ijk}\hat{\lambda}_k$ gives

$$\left(\frac{\partial}{\partial t} + v_j\partial_j + \omega_j\hat{\lambda}_j\right)v_i = \varepsilon_{ijk}\omega_j v_k - \frac{c}{\hbar}\hat{\lambda}_i Q \tag{2.25}$$

which contains a precession-type term in addition to the quantum contribution.

An alternative representation of the internal angular motion is in terms of the velocity fields $v_r(x,\alpha,t)$ conjugate to the Euler angles. These are connected to the components of the vector angular velocity field by the relations

$$\omega_i = \left(A^{-1}\right)_{ir}v_r, \quad v_r = A_{ri}\omega_i, \quad i,r = 1,2,3, \tag{2.26}$$

where

$$A_{ir} = \begin{pmatrix} -\cos\beta & \sin\beta\cot\alpha & -\sin\beta\operatorname{cosec}\alpha \\ \sin\beta & \cos\beta\cot\alpha & -\cos\beta\operatorname{cosec}\alpha \\ 0 & -1 & 0 \end{pmatrix}, \quad \left(A^{-1}\right)_{ir} = \begin{pmatrix} -\cos\beta & 0 & -\sin\alpha\sin\beta \\ \sin\beta & 0 & -\sin\alpha\cos\beta \\ 0 & -1 & -\cos\alpha \end{pmatrix}. \tag{2.27}$$

Relations (2.9) may be written $\hat{\lambda}_i = A_{ir}\partial_r$ and it is easy to show using the result $\left(A^{-1}\right)_{ir}A_{is} = \delta_{rs}$ that $\omega_j\hat{\lambda}_j = v_r\partial_r$. In terms of the conjugate velocities Euler's equations (2.24) and (2.25) become, on substituting (2.26),

$$\left(\frac{\partial}{\partial t} + v_j\partial_j + v_r\partial_r\right)v_s + A_{si}\partial_r\left(A^{-1}\right)_{iq}v_q v_r = -\frac{c}{\hbar}A_{si}\partial_i Q, \quad q,r,s = 1,2,3, \tag{2.28}$$

---

[1] The uniqueness of this identification needs careful discussion. For examination of some of the issues involved in an analogous problem for the Dirac field see [19, 20].

$$\left(\frac{\partial}{\partial t} + v_j \partial_j + v_r \partial_r\right) v_i + \varepsilon_{ijk}\left(A^{-1}\right)_{kr} v_j v_r = -\frac{c}{\hbar}\hat{\lambda}_i Q. \tag{2.29}$$

We shall see in Sec. 4 that the last terms on the left-hand sides of these equations may be attributed a geometrical interpretation.

## C. Fluid paths and Lorentz covariance

The paths $x = x(x_0, \alpha_0, t), \alpha = \alpha(x_0, \alpha_0, t)$ of the fluid particles in $M$ are obtained from the Eulerian velocity functions by solving the differential equations

$$v_i(x, \alpha, t) = \frac{\partial x_i}{\partial t}, \quad v_r(x, \alpha, t) = \frac{\partial \alpha_r}{\partial t}. \tag{2.30}$$

Combining these formulas with (2.23) and (2.26) we have

$$\frac{\partial x_i}{\partial t} = \frac{c}{\hbar} A_{ir} \partial_r S, \quad \frac{\partial \alpha_r}{\partial t} = \frac{c}{\hbar} A_{ri} \partial_i S, \quad i, r = 1, 2, 3, \tag{2.31}$$

where we substitute $x = x(x_0, \alpha_0, t), \alpha = \alpha(x_0, \alpha_0, t)$ on the right-hand sides. These relations generally imply a complex coupling between the translational and angular freedoms. The paths are an analogue in the full wave theory of "rays". One of the achievements of the inverse procedure presented in Sec. 4 - the deduction of the Eulerian Maxwell fields from the paths - is that it demonstrates the consistency of a ray concept in wave optics.

Note that the connection between the fields and the paths embodied in (2.30) breaks relativistic covariance in that quantities having different Lorentz tensorial properties are equated. For example, $v_i$ is not a Lorentz 3-vector (i.e., $\equiv u^i/u^0$ where $u^A$ is a 4-vector)). To see what this means, suppose we repeat the above construction starting in another (primed) inertial frame. Then the new set of trajectories computed from the primed version of (2.30) will not in general be the Lorentz transform of the old set. Whether this symmetry breaking can be removed in an alternative trajectory theory requires further work. It is known in a similar context (of flow lines computed from the ratio of the Poynting vector to the energy density) that covariance can be achieved by introducing extraneous variables [6]. Two points we wish to emphasize, however, are that *a* trajectory path model is possible, and that it is not necessary that this model be covariant in order to deduce the covariant Eulerian field theory.

## D. Classical limit

There are circumstances where $Q$ is both sufficiently small numerically and slowly varying with respect to the (six) coordinates that it and its gradients may be neglected in (2.18), (2.24) and (2.25). We shall term this case the "classical limit" of Maxwell's equations. The motion of a fluid particle can then be expressed as a geodesic for a suitably chosen metric (see (3.9)). In general, these criteria are not the same as the ones that characterize the geometrical optics limit of wave optics [21] (e.g., they involve examination of the angular dependence which is absent in the usual approach) and the limiting trajectories differ from geometric-optic rays (see Secs. 5A and 6A). According to (2.24) and (2.25) along a fluid path the angular velocity is constant and the translational velocity precesses about the angular velocity vector. The deviation from geometrical optics rays is to be expected since the usual energy flow lines are the mean of our paths over the angles. Thus, formula (2.17) may be written

$$\varepsilon_0 c^2 (\mathbf{E} \times \mathbf{B})_i = \int \rho v_i d\Omega \qquad (2.32)$$

and integrating the conservation law (2.22) over the angles the last term drops out and we recover Poynting's theorem relating the quantities (2.16) and (2.17). The limiting flow is a continuous ensemble of non-interacting trajectories in $M$.

## 3. LAGRANGIAN-COORDINATE CONSTRUCTION OF THE WAVEFUNCTION IN A RIEMANNIAN MANIFOLD

### A. Newton's law for a fluid element

We here generalize the method of constructing the wavefunction from hydrodynamic trajectories in Euclidean 3-space presented in [2] to arbitrary coordinates $x^\mu$ in an $N$-dimensional Riemannian manifold $M$ with (static) metric $g_{\mu\nu}(x)$, $\mu, \nu, \dots = 1, \dots, N$. In this space, the history of the fluid is encoded in the positions $\xi(\xi_0, t)$ of the distinct fluid elements at time $t$, each particle being distinguished by its position $\xi_0$ at $t = 0$[2]. We assume that the mapping between these two sets of coordinates is single-valued and differentiable with respect to $\xi_0$ and $t$ to whatever order is necessary, and that the inverse mapping $\xi_0(\xi, t)$ exists and has the same properties.

Let $P_0(\xi_0)$ be the initial density of some continuously distributed quantity in $M$ (mass in ordinary hydrodynamics, energy in our application) and $g = \det g_{\mu\nu}$. Then the quantity in an elementary volume $d^N \xi_0$ attached to the point $\xi_0$ is given by $P_0(\xi_0)\sqrt{-g(\xi_0)}d^N \xi_0$. The conservation of this quantity in the course of the motion of the fluid element is expressed through the relation

$$P(\xi(\xi_0, t))\sqrt{-g(\xi(\xi_0, t))}d^N\xi(\xi_0, t) = P_0(\xi_0)\sqrt{-g(\xi_0)}d^N\xi_0 \qquad (3.1)$$

or

$$P(\xi_0, t) = D^{-1}(\xi_0, t)P_0(\xi_0) \qquad (3.2)$$

where

$$D(\xi_0, t) = \sqrt{g(\xi)/g(\xi_0)}J(\xi_0, t), \quad 0 < D < \infty, \qquad (3.3)$$

and $J$ is the Jacobian of the transformation between the two sets of coordinates:

$$J = \frac{1}{N!}\varepsilon_{\mu_1 \dots \mu_N} \varepsilon^{\nu_1 \dots \nu_N} \frac{\partial \xi^{\mu_1}}{\partial \xi_0^{\nu_1}} \cdots \frac{\partial \xi^{\mu_N}}{\partial \xi_0^{\nu_N}}. \qquad (3.4)$$

We assume that the Lagrangian for the set of fluid particles comprises a kinetic term and an internal potential that represents a certain kind of particle interaction:

---

[2] The preservation of the identity of each fluid element (labelled by its initial position) is of fundamental importance. In the electromagnetic application it apparently constitutes an answer to the objection of Lorentz [22] to the meaningfulness of energy flow lines, which is based upon a claimed loss of identity of individual energy elements when combining with others.

$$L = \int P_0(\xi_0) \left( \tfrac{1}{2} g_{\mu\nu}(\xi) \frac{\partial \xi^\mu}{\partial t} \frac{\partial \xi^\nu}{\partial t} - g^{\mu\nu}(\xi) \frac{c^2 l^2}{8} \frac{1}{P^2} \frac{\partial P}{\partial \xi^\mu} \frac{\partial P}{\partial \xi^\nu} \right) \sqrt{-g(\xi_0)} d^N\xi_0. \tag{3.5}$$

Here $P_0$ and $g_{\mu\nu}$ are prescribed functions, $\xi = \xi(\xi_0, t)$, $l$ is a constant with the dimension of length (introduced, in particular, to ensure that the line element $ds = \sqrt{g_{\mu\nu}(\xi) d\xi^\mu d\xi^\nu}$ has the same dimension; $l$ plays no role in the application of interest), and we substitute for $P$ from (3.2) and write

$$\frac{\partial}{\partial \xi^\mu} = J^{-1} J^\nu_\mu \frac{\partial}{\partial \xi_0^\nu} \tag{3.6}$$

where

$$J^\nu_\mu = \frac{\partial J}{\partial \left( \partial \xi^\mu / \partial \xi_0^\nu \right)} \tag{3.7}$$

is the cofactor of $\partial \xi^\mu / \partial \xi_0^\nu$. The latter satisfies

$$\frac{\partial \xi^\mu}{\partial \xi_0^\nu} J^\sigma_\mu = J \delta^\sigma_\nu. \tag{3.8}$$

It is assumed that $P_0$ and its derivatives vanish at infinity, which ensures that the surface terms in the variational principle vanish.

Varying the coordinates, the Euler-Lagrange equations of motion for the $\xi_0$th fluid particle moving in the "field" of the other particles take the form of Newton's second law in general coordinates:

$$\frac{\partial^2 \xi^\mu}{\partial t^2} + \left\{ \begin{matrix} \mu \\ \nu\sigma \end{matrix} \right\} \frac{\partial \xi^\nu}{\partial t} \frac{\partial \xi^\sigma}{\partial t} = -\frac{cl}{\hbar} g^{\mu\nu} \frac{\partial Q}{\partial \xi^\nu} \tag{3.9}$$

where $\left\{ \begin{matrix} \mu \\ \nu\sigma \end{matrix} \right\} = \tfrac{1}{2} g^{\mu\rho} \left( \partial g_{\sigma\rho} / \partial \xi^\nu + \partial g_{\nu\rho} / \partial \xi^\sigma - \partial g_{\nu\sigma} / \partial \xi^\rho \right)$ and

$$Q = \frac{-\hbar cl}{2\sqrt{-gP}} \frac{\partial}{\partial \xi^\mu} \left( \sqrt{-g} \, g^{\mu\nu} \frac{\partial \sqrt{P}}{\partial \xi^\nu} \right). \tag{3.10}$$

Here we have written the force term on the right-hand side of (3.9) in condensed form and substituting for $P$ from (3.2) and for the derivatives with respect to $\xi$ from (3.6) we obtain a highly complex fourth order (in $\xi_0$) local nonlinear partial differential equation. We shall see that from the solutions $\xi = \xi(\xi_0, t)$, subject to specification of $\partial \xi_0^\mu / \partial t$ whose determination is discussed next, we may derive solutions to Schrödinger's equation.

## B. Quasi-potential flow

To obtain a flow that is representative of Schrödinger evolution we need to restrict the initial conditions of (3.9) to those that correspond to what we term "quasi-potential" flow. This means that the initial covariant components of the velocity field are of the form (we introduce the factor $cl/\hbar$ with an eye to the electromagnetic application)

$$g_{\mu\nu}(\xi_0)\frac{\partial \xi_0^\mu}{\partial t} = \frac{cl}{\hbar}\frac{\partial S_0(\xi_0)}{\partial \xi_0^\nu} \tag{3.11}$$

but the flow is not irrotational everywhere because the potential $S_0(\xi_0)$ obeys the quantization condition

$$\oint_C \frac{\partial S_0(\xi_0)}{\partial \xi_0^\mu} d\xi_0^\mu = nh, \quad n \in Z, \tag{3.12}$$

where $C$ is a closed curve composed of fluid particles. The requirement (3.12) evidently restricts the circulation of the covariant components of the initial velocity (3.11). If it exists, the vorticity occurs in nodal regions (where the density vanishes) and it is assumed that $C$ passes through a region of "good" fluid, where $P_0 \neq 0$. To show that these assumptions imply motion characteristic of Schrödinger evolution we first demonstrate that they are preserved by the dynamical equation. To this end, we use the method based on Weber's transformation [23, 24] together with an analogue of Kelvin's circulation theorem.

We first multiply (3.9) by $g_{\sigma\mu}\partial\xi^\sigma/\partial\xi_0^\rho$ and integrate between the time limits $(0,t)$. The term involving the Christoffel symbols drops out and we obtain

$$g_{\sigma\mu}\big(\xi(\xi_0,t)\big)\frac{\partial \xi^\sigma}{\partial \xi_0^\rho}\frac{\partial \xi^\mu}{\partial t} = g_{\rho\mu}(\xi_0)\frac{\partial \xi_0^\mu}{\partial t} + \frac{\partial}{\partial \xi_0^\rho}\int_0^t\left(\tfrac{1}{2}g_{\mu\nu}\big(\xi(\xi_0,t)\big)\frac{\partial \xi^\mu}{\partial t}\frac{\partial \xi^\nu}{\partial t} - \frac{cl}{\hbar}Q\right)dt. \tag{3.13}$$

Then, substituting (3.11),

$$g_{\sigma\mu}\frac{\partial \xi^\sigma}{\partial \xi_0^\rho}\frac{\partial \xi^\mu}{\partial t} = \frac{cl}{\hbar}\frac{\partial S}{\partial \xi_0^\mu}, \quad S = S_0 + \int_0^t\left(\tfrac{1}{2}\frac{\hbar}{cl}g_{\mu\nu}\frac{\partial \xi^\mu}{\partial t}\frac{\partial \xi^\nu}{\partial t} - Q\right)dt. \tag{3.14}$$

The left-hand side of (3.14) gives the covariant velocity components at time $t$ with respect to the $\xi_0$-coordinates and these obviously form a gradient vector. To obtain the $\xi$-components we multiply by $J^{-1}J_\nu^\rho$ and use (3.6) and (3.8) to get

$$g_{\mu\nu}\frac{\partial \xi^\nu}{\partial t} = \frac{cl}{\hbar}\frac{\partial S}{\partial \xi^\mu} \tag{3.15}$$

where $S = S\big(\xi_0(\xi,t),t\big)$. Thus, for all time the covariant velocity of each particle is the gradient of a potential with respect to the current position.

To complete the demonstration, we note that the motion is quasi-potential since the system satisfies (3.9) and hence possesses an acceleration potential (proportional to $Q$). It follows that the value (3.12) of the circulation is preserved following the flow:

$$\frac{\partial}{\partial t}\oint_{C(t)}g_{\mu\nu}\frac{\partial \xi^\nu}{\partial t}d\xi^\mu = 0 \tag{3.16}$$

where $C(t)$ is the evolute of the fluid particles that compose $C$. We conclude that each particle retains forever the quasi-potential property if it possesses it at any moment.

## C. Derivation of Schrödinger's equation

The fundamental link between the particle (Lagrangian) and wave-mechanical (Eulerian) pictures is defined by the following expression for the Eulerian density:

$$P(x,t)\sqrt{-g(x)} = \int \delta\big(x - \xi(\xi_0,t)\big) P_0(\xi_0) \sqrt{-g(\xi_0)}\, d^N\xi_0. \tag{3.17}$$

The corresponding formula for the Eulerian velocity is contained in the expression for the current:

$$P(x,t)\sqrt{-g(x)}\, v^\mu(x,t) = \int \frac{\partial \xi^\mu(\xi_0,t)}{\partial t} \delta\big(x - \xi(\xi_0,t)\big) P_0(\xi_0) \sqrt{-g(\xi_0)}\, d^N\xi_0. \tag{3.18}$$

Evaluating the integrals, (3.17) and (3.18) are equivalent to the following local expressions

$$P(x,t)\sqrt{-g(x)} = J^{-1}\Big|_{\xi_0(x,t)} P_0\big(\xi_0(x,t)\big)\sqrt{-g\big(\xi_0(x,t)\big)} \tag{3.19}$$

$$v^\mu(x,t) = \frac{\partial \xi^\mu(\xi_0,t)}{\partial t}\bigg|_{\xi_0(x,t)}. \tag{3.20}$$

These formulas enable us to translate the Lagrangian flow equations into Eulerian language. Differentiating (3.17) with respect to $t$ and using (3.18) we deduce the continuity equation

$$\frac{\partial P}{\partial t} + \frac{1}{\sqrt{-g}} \frac{\partial}{\partial x^\mu}\big(P\sqrt{-g}\, v^\mu\big) = 0. \tag{3.21}$$

Next, differentiating (3.18) and using (3.9) and (3.21) we get the analogue of Euler's classical equation:

$$\frac{\partial v^\mu}{\partial t} + v^\nu \frac{\partial v^\mu}{\partial x^\nu} + \left\{ \begin{matrix} \mu \\ \nu\sigma \end{matrix} \right\} v^\nu v^\sigma = -\frac{cl}{\hbar} g^{\mu\nu} \frac{\partial Q}{\partial x^\nu}, \tag{3.22}$$

where $Q$ is given by (3.10) with $\xi$ replaced by $x$. Finally, the quasi-potential condition (3.15) becomes

$$v^\mu = \frac{cl}{\hbar} g^{\mu\nu} \frac{\partial S(x,t)}{\partial x^\nu}. \tag{3.23}$$

Formulas (3.19) and (3.20) give the general solution of the coupled continuity and Euler equations (3.21) and (3.22) in terms of the paths and initial density.

To establish the connection between the Eulerian equations and Schrödinger's equation we note that, using (3.23), (3.22) can be written

$$\frac{\partial}{\partial x^\mu}\left( \frac{\partial S}{\partial t} + \frac{1}{2} \frac{cl}{\hbar} g^{\nu\sigma} \frac{\partial S}{\partial x^\nu} \frac{\partial S}{\partial x^\sigma} + Q \right) = 0. \tag{3.24}$$

The quantity in brackets is thus a function of time. Since the addition of a function of time to $S$ does not affect the velocity field, we may absorb the function in $S$, i.e., set it to zero. Then

$$\frac{\partial S}{\partial t} + \frac{1}{2}\frac{cl}{\hbar}g^{\nu\sigma}\frac{\partial S}{\partial x^{\nu}}\frac{\partial S}{\partial x^{\sigma}} + Q = 0. \tag{3.25}$$

Combining (3.25) with (3.21) (where we substitute (3.23)) we find that the function $\psi(x,t) = \sqrt{P}\exp(iS/\hbar)$ obeys the free Schrödinger equation in general coordinates:

$$i\hbar\frac{\partial\psi}{\partial t} = \frac{-\hbar cl}{2\sqrt{-g}}\frac{\partial}{\partial x^{\mu}}\left(\sqrt{-g}\,g^{\mu\nu}\frac{\partial\psi}{\partial x^{\nu}}\right) \tag{3.26}$$

(for a system of "mass" $\hbar/cl$). We have deduced the wave equation from the collective particle motion obeying the Lagrangian path equation (3.9) subject to the quasi-potential requirement. The quantization condition (3.16) becomes here

$$\oint_{C_0}\frac{\partial S(x,t)}{\partial x^{\mu}}dx^{\mu} = nh, \quad n \in Z, \tag{3.27}$$

where $C_0$ is a closed curve fixed in space that does not pass through nodes. This is a consistent subsidiary condition on solutions since it is easy to see using (3.24) that the value of (3.27) is preserved in time as long as nodes do not cross $C_0$.

## 4. DEDUCTION OF THE ELECTROMAGNETIC FIELD FROM THE TRAJECTORIES

We specialize the treatment of the last section to the manifold $M = \Re^3 \otimes SO(3)$ with coordinates $x^{\mu} = (x_i, \alpha_r)$, metric

$$g^{\mu\nu} = \begin{pmatrix} 0 & l^{-1}A_{ir} \\ l^{-1}A_{ri} & 0 \end{pmatrix}, \quad g_{\mu\nu} = \begin{pmatrix} 0 & l(A^{-1})_{ir} \\ l(A^{-1})_{ri} & 0 \end{pmatrix} \quad i,r = 1,2,3, \tag{4.1}$$

where $A_{ir}$ is given by (2.27), and density $P = \rho/l^3$. Using, in particular, the results $\partial_r(\sqrt{-g}\,g^{ir}) = 0$ and $lg^{ir}\partial_r = \hat{\lambda}_i$, and inserting the latter in the relations $\left[\hat{\lambda}_i, \hat{\lambda}_j\right] = -\varepsilon_{ijk}\hat{\lambda}_k$ which gives $g^{ir}(\partial_s g_{rj} - \partial_j g_{sj}) = l^{-1}\varepsilon_{ijk}g_{sk}$, we find that the gradient relation (3.23) becomes the electromagnetic Eulerian relations (2.23), the conservation equation (3.21) becomes (2.19), Euler's equation (3.22) becomes (2.28) and (2.29), the quantum potential (3.10) (with $\xi$ replaced by $x$) becomes (2.20), the Hamilton-Jacobi-like equation (3.25) becomes (2.18), the Schrödinger equation (3.26) becomes Maxwell's equations (2.11), and the circulation condition (3.27) becomes (2.21).

The results of the previous section therefore provide us with a continuum mechanics model from which we may deduce Maxwell's equations as the Eulerian counterpart to the equations of motion of the Lagrangian trajectories, and in particular with an algorithm to compute the electromagnetic field from the latter. Writing $\xi^{\mu} = (q_i, \theta_r)$ for the Lagrangian coordinates, the Lagrangian (3.5) becomes

$$L = \int l\rho_0(q_0,\theta_0)\left(\left(A^{-1}\right)_{ir}\frac{\partial q_i}{\partial t}\frac{\partial \theta_r}{\partial t} - A_{ir}\frac{c^2}{4}\frac{1}{\rho^2}\frac{\partial \rho}{\partial q_i}\frac{\partial \rho}{\partial \theta_r}\right)\sin\theta_{01}d^3\theta_0 d^3q_0. \tag{4.2}$$

Newton's law (3.9) reduces to the coupled relations

$$\frac{\partial^2 q_i}{\partial t^2} + \varepsilon_{ijk}\left(A^{-1}\right)_{kr}\frac{\partial q_j}{\partial t}\frac{\partial \theta_r}{\partial t} = -\frac{c}{\hbar}A_{ir}\frac{\partial Q}{\partial \theta_r} \tag{4.3}$$

$$\frac{\partial^2 \theta_s}{\partial t^2} + A_{si}\frac{\partial}{\partial \theta_r}\left(A^{-1}\right)_{iq}\frac{\partial \theta_q}{\partial t}\frac{\partial \theta_r}{\partial t} = -\frac{c}{\hbar}A_{si}\frac{\partial Q}{\partial q_i} \tag{4.4}$$

where $A_{ir}$ is given by (2.27) with $\alpha_r$ replaced by $\theta_r(q_0,\theta_0,t)$ and we substitute

$$\rho(q_0,\theta_0,t) = D^{-1}(q_0,\theta_0,t)\rho_0(q_0,\theta_0) \tag{4.5}$$

into

$$Q = -c\hbar A_{ir}\frac{1}{\sqrt{\rho}}\frac{\partial^2 \sqrt{\rho}}{\partial \theta_r \partial q_i} \tag{4.6}$$

with

$$\frac{\partial}{\partial q_i} = J^{-1}\left(J_{ij}\frac{\partial}{\partial q_{0j}} + J_{is}\frac{\partial}{\partial \theta_{0s}}\right), \quad \frac{\partial}{\partial \theta_r} = J^{-1}\left(J_{rj}\frac{\partial}{\partial q_{0j}} + J_{rs}\frac{\partial}{\partial \theta_{0s}}\right) \tag{4.7}$$

Given the initial wavefunction $\psi_0(x,\alpha) = G_{0a}(x)u_a(\alpha) = \sqrt{\rho_0}\exp(iS_0/\hbar)$ we can compute the wavefunction for all $x,\alpha,t$, up to a global phase, as follows. First, solve (4.3) and (4.4) subject to the initial conditions $\partial q_{0i}/\partial t = (c/\hbar)A_{ir}(\theta_0)\partial S_0/\partial \theta_{0r}$ and $\partial \theta_{0r}/\partial t = (c/\hbar)A_{ri}(\theta_0)\partial S_0/\partial q_{0i}$ to get the set of trajectories for all $q_0,\theta_0,t$. Next, invert these functions and substitute $q_0(x,\alpha,t)$ and $\theta_0(x,\alpha,t)$ in the right-hand side of (4.5) to find $\rho(x,\alpha,t)$ and in the right-hand sides of the equations

$$\partial_r S = \frac{\hbar}{c}\left(A^{-1}\right)_{ir}\frac{\partial q_i}{\partial t}, \quad \partial_i S = \frac{\hbar}{c}\left(A^{-1}\right)_{ri}\frac{\partial \theta_r}{\partial t}, \quad i,r = 1,2,3, \tag{4.8}$$

to get $S$ up to an additive function of time, $\hbar f(t)$. To fix this function, apart from an additive constant, use (2.18). We obtain then the following formula for the wavefunction:

$$\psi(x,\alpha,t) = \sqrt{\left(D^{-1}\rho_0\right)\Big|_{\substack{q_0(x,\alpha,t)\\ \theta_0(x,\alpha,t)}}}$$
$$\times \exp\left[\frac{i}{c}\int \left(A^{-1}\right)_{ir}\partial q_i/\partial t\Big|_{\substack{q_0(x,\alpha,t)\\ \theta_0(x,\alpha,t)}}d\alpha_r + \left(A^{-1}\right)_{ri}\partial \theta_r/\partial t\Big|_{\substack{q_0(x,\alpha,t)\\ \theta_0(x,\alpha,t)}}dx_i + if(t)\right]. \tag{4.9}$$

Finally, the components of the time-dependent electromagnetic field may be read off from the formula (2.7) where (inverse of (2.12))

$$G_a = \int \psi(x,\alpha) u_a^*(\alpha) d\Omega. \tag{4.10}$$

Note that the trajectories depend on the basis set $u_a(\alpha)$ (which enter the calculation through the initial wavefunction) and would be different if a different choice to (2.13) were made. This will not affect the field values found from (4.10).

## 5. EULERIAN PROPAGATION

### A. Maxwell's equations as a first quantized theory

The alternative Eulerian hydrodynamic picture we have developed offers the possibility of deriving a novel expression for the propagator $K$ of the electromagnetic field. As an aid to computing $K$, we shall invoke our demonstration that Maxwell's equations in flat space can be expressed as the Schrödinger equation for a massive system in a certain curved six-space (namely, (3.26) with metric (4.1)). Then the desired function can be determined from the known formula for the curved-space short-time propagator in terms of the classical action. In our case, the "classical" theory for which the action is to be computed is the limiting case of the Maxwellian "quantum" theory described in Sec. 2D. Note that the relevant trajectories to be employed in evaluating the action are the classical Lagrangian paths and not, say, the rays of geometrical optics. We first confirm that we have correctly identified the classical theory by showing how Maxwell's equations may be obtained from it via canonical quantization.

The classical system whose canonical quantization yields (3.26) has Lagrangian $L = (\hbar/2cl) g_{\mu\nu} \dot{x}^\mu \dot{x}^\nu$ with $\dot{x}^\mu = dx^\mu/dt$. Inserting the metric (4.1), the Lagrangian becomes

$$L(x,\alpha,\dot{x},\dot{\alpha}) = (\hbar/c)(A^{-1})_{ir} \dot{x}_i \dot{\alpha}_r = (\hbar/c) \dot{x}_i \omega_i, \tag{5.1}$$

using (2.26). The system, regarded as a particle with an attached frame defining an orientation, traverses an orbit in the configuration space $M = \Re^3 \otimes SO(3)$ labelled by the Cartesian coordinates $x_i(t)$ and Euler angles $\alpha_r(t)$. Note that the factor $\hbar/c$, which comes from the mass parameter $\hbar/cl$, is crucial to obtaining the correct quantum theory but plays no role in the classical dynamics. Varying the six independent coordinates, the Euler-Lagrange equations are

$$\delta x_i: \quad \ddot{\alpha}_s + A_{si} \frac{\partial}{\partial \alpha_r}(A^{-1})_{iq} \dot{\alpha}_q \dot{\alpha}_r = 0 \tag{5.2}$$

$$\delta \alpha_r: \quad \ddot{x}_i - \varepsilon_{ijk}(A^{-1})_{jr} \dot{x}_k \dot{\alpha}_r = 0. \tag{5.3}$$

These relations may be written

$$\dot{\omega}_i = 0, \quad \ddot{x}_i = \varepsilon_{ijk} \omega_j \dot{x}_k \tag{5.4}$$

and the space orbit is

$$x_i(t) = \frac{1}{\omega^2} \varepsilon_{ijk} \omega_j \dot{x}_{0k}(1 - \cos\omega t) + \frac{1}{\omega}\left(\dot{x}_{0i} - \frac{\dot{x}_{0j}\omega_j}{\omega^2}\omega_i\right)\sin\omega t + \frac{\dot{x}_{0j}\omega_j}{\omega^2}\omega_i t + x_{0i}, \tag{5.5}$$

where $\omega = |\boldsymbol{\omega}|$. The path is a helix about the constant vector $\omega_i$ and the body spins about this axis with uniform angular speed $\omega$. Defining the canonical momenta to be

$$p_i = \frac{\partial L}{\partial \dot{x}_i} = (\hbar/c)\omega_i, \quad \pi_r = \frac{\partial L}{\partial \dot{\alpha}_r} = (\hbar/c)(A^{-1})_{ir}\dot{x}_i, \tag{5.6}$$

the Hamiltonian is

$$H(x,\alpha,p,\pi) = p_i\dot{x}_i + \pi_r\dot{\alpha}_r - L = (\hbar/c)^{-1}A_{ir}\pi_r p_i, \tag{5.7}$$

using $(A^{-1})_{ir}A_{is} = \delta_{rs}$.

There are three components to the quantization scheme that results in Maxwell's equations. The first is to replace the canonical variables by operators, and Poisson brackets by commutators, following the usual procedure of the wave mechanics of rotators [6]. Then

$$p_i \to \hat{p}_i = -i\hbar\partial/\partial x_i, \quad \pi_r \to \hat{\pi}_r = -i\hbar\partial/\partial\alpha_r, \quad H \to \hat{H} = (\hbar/c)^{-1}A_{ir}\hat{\pi}_r\hat{p}_i = -ic\hat{\lambda}_i\hat{p}_i, \tag{5.8}$$

(making a specific choice of operator ordering) and we obtain the Schrödinger equation (2.11). At this stage the wavefunction on $M$ is a superposition of integer-spin states.

We next observe that the Schrödinger equation splits up into a set of independent wave equations for each spin. Restricting $\psi$ to be a spin 1 simultaneous eigenstate of the total angular momentum operator $\hat{\mathbf{M}}^2$ and the anomalous angular momentum operator $\hat{M}'_3$ [6] constitutes the second aspect of the quantization process.

Finally, we note that in setting up the classical model we did not worry about relativistic covariance and indeed the quantization procedure generates only the pair of field equations (2.1), a non-covariant fragment of Maxwell's equations. To obtain the full covariant set of field equations we must specify the correct initial conditions stated in Sec. 2A so that equations (2.1) imply the remaining two equations (2.2). These initial conditions are the third ingredient of the quantization process. This state of affairs is reminiscent of Feynman's derivation of Maxwell's equations from Newton's force law subject to the position-momentum commutation relations [25]. In fact, what is obtained by Feynman's procedure is a (different) non-covariant portion of Maxwell's equations, the complementary equations that produce the full Lorentz covariant set being introduced as definitions of the sources [26].

## B. Eulerian propagator

The short-time propagator solution to the curved-space Schrödinger equation (3.26) generating evolution from $x_0^\mu$ to $x^\mu$ in time $t$ is given by [27][3]

$$K(x,t;x_0,0) = (2\pi i\hbar)^{-N/2}(-g(x))^{-1/4}(-g(x_0))^{-1/4}\sqrt{|d|}\exp(i\delta)\exp(ilcRt/12)\exp(iW/\hbar) \tag{5.9}$$

where $W(x,t;x_0,0) = \int_{x_0,0}^{x,t} L\,dt$ is the classical action,

$$d(x,t;x_0,0) = \det\left(-\frac{\partial^2 W}{\partial x^\mu \partial x_0^\nu}\right) \tag{5.10}$$

---

[3] For further discussion on the curved-space propagator see [28].

is the van Vleck determinant, $R$ is the curvature scalar, and $\delta = 0\,(\pi/2)$ for $d > 0\,(<0)$. This function has the property $\lim_{t \to 0} K(x,t;x_0,0) = (-g)^{-1/2}\delta(x-x_0)$. For $t \neq 0$, $K$ obeys exactly the classical wave equation (comprising the classical Hamilton-Jacobi equation and the continuity equation) [6],

$$i\hbar \frac{\partial K}{\partial t} = \frac{-\hbar cl}{2\sqrt{-g}} \frac{\partial}{\partial x^\mu}\left(\sqrt{-g}\,g^{\mu\nu}\frac{\partial K}{\partial x^\nu}\right) - QK, \tag{5.11}$$

where $Q$ is the quantum potential (2.20) constructed from $|K|$. In the limit $t \to 0$, $K$ obeys the Schrödinger equation.

In our application the curvature scalar derived from the metric (4.1) is zero. Using the Lagrangian (5.1) the action function is

$$W(x,\alpha,t;x_0,\alpha_0,0) = (\hbar/c)\int_{x_0,\alpha_0,0}^{x,\alpha,t} \dot{x}_i \omega_i\, dt. \tag{5.12}$$

Differentiating (5.5) we have $\dot{x}_i\omega_i = \dot{x}_{0i}\omega_i$ and so $W = (\hbar/c)\dot{x}_{0i}\omega_i t$. Again from (5.5), $x_i(t)\omega_i = \dot{x}_{0j}\omega_j t + x_{0i}\omega_i$. The action therefore becomes

$$W(x,\alpha,t;x_0,\alpha_0,0) = (\hbar/c)(x_i - x_{0i})\omega_i(\alpha,t;\alpha_0,0). \tag{5.13}$$

It remains to express $\omega_i$ in terms of the initial and final angles and the time. For this, we adapt a method used to obtain the action for a symmetric top [29]. We express the orientation of the system at time $t$ by a matrix $U(\alpha,\beta,\gamma) \in SU(2)$:

$$U(\alpha,\beta,\gamma) = \begin{pmatrix} e^{i(\gamma+\beta)/2}\cos(\alpha/2) & ie^{i(\beta-\gamma)/2}\sin(\alpha/2) \\ ie^{i(\gamma-\beta)/2}\sin(\alpha/2) & e^{-i(\gamma+\beta)/2}\cos(\alpha/2) \end{pmatrix} \tag{5.14}$$

The relation between the initial and final orientations may then be written as $U(\alpha,\beta,\gamma) = \pm\exp(-i\Phi\hat{n}_i\sigma_i/2)U_0(\alpha_0,\beta_0,\gamma_0)$ where $\hat{n}_i = \omega_i/\omega$ is a unit vector defining the axis of rotation, $\Phi = \omega t$ is the angle of rotation, and $\sigma_i$ are the Pauli matrices. This formula gives $\exp(-i\Phi\hat{n}_i\sigma_i/2) = \pm U U_0^{-1}$ and hence

$$\cos(\Phi/2) = \pm\tfrac{1}{2}\text{Tr}\left(U U_0^{-1}\right), \quad \sin(\Phi/2)\hat{n}_i = \pm(i/2)\text{Tr}\left(U U_0^{-1}\sigma_i\right). \tag{5.15}$$

The first formula in (5.15) gives $\omega$ as a function of the angles and $t$ and the second formula, on using the first, gives $\hat{n}_i$. Combining these results we obtain

$$\omega_i(\alpha,t;\alpha_0,0) = \frac{\pm i\text{Tr}\left(U U_0^{-1}\sigma_i\right)\cos^{-1}\left[\tfrac{1}{2}\text{Tr}\left(U U_0^{-1}\right) + n\pi\right]}{t\sqrt{1 - \left(\tfrac{1}{2}\text{Tr}\left(U U_0^{-1}\right)\right)^2}}, \tag{5.16}$$

where the + (-) sign corresponds to $n$ even (odd). This establishes the time dependence of $\omega_i$ and we shall write $\omega_i(\alpha,t;\alpha_0,0) = \kappa_i(\alpha,\alpha_0)/t$. The computation of the propagator is completed by evaluating the amplitude. We have $(-g(\alpha))^{-1/4}(-g(\alpha_0))^{-1/4} = l^{-3}(\sin\alpha\sin\alpha_0)^{-1/2}$ and inserting (5.13) in (5.10) gives

$$d(x,\alpha,t;x_0,\alpha_0,0) = (\hbar/ct)^6 \det\left(\frac{\partial\kappa_i}{\partial\alpha_r}\right)\det\left(\frac{\partial\kappa_i}{\partial\alpha_{0r}}\right). \tag{5.17}$$

The full expression for the propagator of the electromagnetic field in the Eulerian picture is therefore:

$$K(x,\alpha,t;x_0,\alpha_0,0) = (2\pi i ctl)^{-3}(\sin\alpha_0\sin\alpha)^{-1/2}\sqrt{\left|\det\left(\frac{\partial\kappa_i}{\partial\alpha_r}\right)\det\left(\frac{\partial\kappa_i}{\partial\alpha_{0r}}\right)\right|}$$
$$\times \exp(i\delta)\exp\left[i(x_i - x_{0i})\kappa_i(\alpha,\alpha_0)/ct\right]. \tag{5.18}$$

An important property of $K$ is that the amplitude is independent of $x_i$ and hence the quantum potential (2.20) for this function vanishes. The classical wave equation (5.11) therefore coincides exactly with the Schrödinger equation (2.11) and $K$ is the exact propagator. This happens even though the classical Hamiltonian (5.7) is not quadratic in the phase space variables.

The Eulerian counterpart to the propagation equation (4.9) of the Lagrangian picture is

$$\psi(x,\alpha,t) = \int K(x,\alpha,t;x_0,\alpha_0,0)\psi_0(x_0,\alpha_0)l^3 d\Omega_0 d^3x_0. \tag{5.19}$$

Together these evolution equations provide complementary – trajectory and field-theoretic – descriptions for the propagation of the electromagnetic field. $K$ is, of course, independent of $\hbar$ and its $l$-dependence cancels that of the measure.

## 6. COMPLEMENTARY PICTURES OF THE PROPAGATION OF LIGHT

### A. Construction by trajectories

Using the method of Sec. 4, we aim to compute the time dependence of the electromagnetic field whose initial form is

$$E_{0i} = (E\cos kz, 0, 0), \quad B_{0i} = (0, (1/c)E\cos kz, 0). \tag{6.1}$$

With this choice the corresponding initial wavefunction is $\psi_0 = G_{01}u_1$, or

$$\psi_0(q_0,\theta_0) = -\left(\sqrt{3}/2\sqrt{2}\pi\right)E\cos kq_{03}\sin\theta_{01}e^{-i\theta_{02}}. \tag{6.2}$$

Between nodes the initial phase is, up to an additive constant,

$$S_0 = -\hbar\theta_{02} + \eta_1(\theta_{01}) + \eta_2(q_{03}) + \pi, \quad \eta_1 = \begin{cases} 0, & 0 < \theta_{01} < \pi/2 \\ \pi\hbar, & \pi/2 < \theta_{01} < \pi \end{cases}, \quad \eta_2 = \begin{cases} 0, & \cos kq_{03} > 0 \\ \pi\hbar, & \cos kq_{03} < 0 \end{cases}. \tag{6.3}$$

The initial velocity components are therefore

$$\partial q_{0i}(q_0,\theta_0)/\partial t = c(-\sin\theta_{02}\cot\theta_{01}, -\cos\theta_{02}\cot\theta_{01}, 1), \quad \partial\theta_{0r}(q_0,\theta_0)/\partial t = 0. \tag{6.4}$$

To simplify matters, we shall seek solutions to (4.3) and (4.4) that generate a time-dependent wavefunction whose spatial dependence is on $z$ alone. The Hamiltonian in the Schrödinger equation (2.10) then reduces to $-ic\hat{M}_3\partial_3\psi(x,\alpha)$, which preserves the spin-dependence of $\psi_0$. A consequence is that, since $\rho$ is independent of $\theta_3$, the quantum potential (4.6) vanishes. Combining (2.24) evaluated along an orbit with (6.4) then implies that the angular velocity vector is zero. This implies via (2.26) that $\partial\theta_r/\partial t = 0$. Substituting in (4.3) then gives $\partial^2 q_i/\partial t^2 = 0$ and hence the space trajectories are uniform and rectilinear (corresponding to the $\omega_i \to 0$ limit of (5.5)):

$$q_i(q_0,\theta_0,t) = V_i t + q_{0i}, \quad \theta_r(q_0,\theta_0,t) = \theta_{0r},$$
$$V_i = c\left(-\sin\theta_{02}\cot\theta_{01}, -\cos\theta_{02}\cot\theta_{01}, 1\right). \tag{6.5}$$

This solution gives $D = 1$ and so, substituting in (4.5) and (4.8), we get

$$\rho(x,\alpha,t) = \rho_0(q_0,\alpha), \quad \partial_r S = \hbar(0,-1,0), \quad \partial_i S = 0,$$
$$q_{0i} = x_i - ct\left(-\sin\alpha_2\cot\alpha_1, -\cos\alpha_2\cot\alpha_1, 1\right). \tag{6.6}$$

Hence, since from (2.18) $f(t) = $ constant,

$$\rho(x,\alpha,t) = \left(\sqrt{3}/2\sqrt{2}\pi\right)^2 E^2\cos^2 k(z-ct)\sin^2\alpha_1, \quad S(x,\alpha,t) = -\hbar\beta + \eta_1(\alpha) + \eta_2(z,t) + \pi,$$
$$\eta_1 = \begin{cases} 0, & 0 < \alpha < \pi/2 \\ \pi\hbar, & \pi/2 < \alpha < \pi \end{cases}, \quad \eta_2 = \begin{cases} 0, & \cos k(z-ct) > 0 \\ \pi\hbar, & \cos k(z-ct) < 0 \end{cases}, \tag{6.7}$$

and the corresponding wavefunction is

$$\psi(x,\alpha,t) = -\left(\sqrt{3}/2\sqrt{2}\pi\right)E\cos k(z-ct)\sin\alpha e^{-i\beta}. \tag{6.8}$$

The implied electric and magnetic fields are

$$E_i = \left(E\cos k(ct-z), 0, 0\right), \quad B_i = \left(0, (1/c)E\cos k(ct-z), 0\right). \tag{6.9}$$

They represent a plane wave propagating in the $z$-direction, which is the correct solution to Maxwell's equations corresponding to the initial conditions (6.1).

Note that we obtain oscillatory behaviour of the Eulerian variables from a model in which the individual fluid elements do not oscillate. We thus circumvent one of the problems with nineteenth century ether models where it was considered that the elements of a continuum must vibrate in order to support wave motion, a requirement that was difficult to reconcile with electromagnetic phenomena [30] (a detailed comparison between the present model and the older ether models will be presented elsewhere). Varying the initial angular coordinates, the projected flow in $\Re^3$ is a complex crosshatch of lines in the $q_1 q_2$-plane with common motion in the $q_3$-direction. The speed of each element, $|\mathbf{V}| = |c\cosec\theta_{01}|$, obeys $c \leq |\mathbf{V}| < \infty$. One might regard the occurrence of superluminal speeds as evidence that the Lagrangian model should be afforded the status of only a mathematical tool, analogous to, say, the decomposition of the wavefunction in the path-integral formalism. It augments other well-known instances of superluminality in electromagnetic theory (the significance of which continues to be debated (e.g., [31, 32])). Performing a weighted sum of the velocity over the angles to get the Poynting vector (2.32), the collective $x$- and $y$-motions respectively cancel to give the conventional geometrical optics rays propagating at speed $c$ in the $z$-direction.

**B. Construction by propagator**

The point-to-point Lagrangian description of the evolution of an electromagnetic wave just given may be contrasted with the many-to-many Eulerian mapping embodied in the formula (5.19). Applying the latter to the initial wavefunction (6.2) we obtain

$$\psi(x,\alpha,t) = -\left(\sqrt{3}/2\sqrt{2}\pi\right)E(2\pi ict)^{-3}(\sin\alpha)^{-1/2}$$

$$\times \int \sqrt{\left|\det\left(\frac{\partial\kappa_i}{\partial\alpha_r}\right)\det\left(\frac{\partial\kappa_i}{\partial\alpha_{0r}}\right)\right|}\, e^{i\delta}e^{i(x_i-x_{0i})\kappa_i/ct}\cos k z_0 \sin^{3/2}\alpha_0 e^{-i\beta_0}d\alpha_0 d\beta_0 d\gamma_0 d^3 x_0 \quad (6.10)$$

The integrations with respect to $x_0$, $y_0$ and $z_0$ produce the terms $\frac{1}{2}(2\pi)^3\delta(\kappa_1/ct)\delta(\kappa_2/ct)\delta(\kappa_3/ct\mp k)$. From the second formula in (2.26), the relations $\kappa_1=\kappa_2=0, \kappa_3=\pm kct$ imply $\alpha_0=\alpha, \gamma_0=\gamma, \beta_0=\beta\mp kct$ and $x_i\kappa_i/ct=z(\beta_0-\beta)/ct$. Expressing the triple product of delta functions in terms of delta functions of the angles we obtain

$$\psi(x,\alpha,t) = -\frac{1}{2}\left(\sqrt{3}/2\sqrt{2}\pi\right)E(ict)^{-3}(\sin\alpha)^{-1/2}\int\sqrt{\left|\det\left(\frac{\partial\kappa_i}{\partial\alpha_r}\right)\det\left(\frac{\partial\kappa_i}{\partial\alpha_{0r}}\right)\right|}\, e^{i\delta}e^{iz(\beta_0-\beta)/ct}\sin^{3/2}\alpha_0 e^{-i\beta_0}$$

$$\times\left\{\left[\det\left(\frac{\partial}{\partial\alpha_{0r}}\left(\frac{\kappa_i}{ct}\right)\right)\right]^{-1}_{\substack{\alpha_0=\alpha\\\beta_0=\beta-kct\\\gamma_0=\gamma}}\delta(\beta_0-\beta+kct)+\left[\det\left(\frac{\partial}{\partial\alpha_{0r}}\left(\frac{\kappa_i}{ct}\right)\right)\right]^{-1}_{\substack{\alpha_0=\alpha\\\beta_0=\beta+kct\\\gamma_0=\gamma}}\delta(\beta_0-\beta-kct)\right\} \quad (6.11)$$

$$\times\delta(\alpha_0-\alpha)\delta(\gamma_0-\gamma)d\alpha_0 d\beta_0 d\gamma_0$$

Noting that the first determinant under the square root is the negative of the second ($\delta=\pi/2$), the determinants in (6.11) drop out and we obtain (6.8).

# 7. CONCLUSION

We have established that electromagnetic phenomena, conventionally described only in a field-theoretic language, admit a complementary description in terms of a (continuously-) many- "particle" system possessing an interaction potential of a certain kind. The approach may be regarded as providing a formal statement of "wave-particle" duality, and entails the derivation of Maxwell's equations – regarded as Eulerian fluid equations - from the Lagrangian continuum model.

   We may treat this derivation as a method of "quantization" having the following characteristics: starting from a single particle having translational and orientational freedoms (and Lagrangian (5.1)), pass to a continuum of particles (with Lagrangian given by the first term in (4.2)), introduce an interparticle interaction (the second term in (4.2)), and specify appropriate initial conditions. The method attributes a fundamental formal significance to the quantum potential energy, beyond its original purely interpretational aspect [6]. In this approach the general solution to Maxwell's equations is expressed in terms of a congruence of paths (with due reference to the angular freedoms). This technique of quantization is to be distinguished from conventional canonical quantization, which we have also employed to derive Maxwell's equations starting from the single-particle classical model. An alternative expression for the general solution has been developed using the propagator, which has been computed from the classical paths. It is anticipated that the exact scheme for computing the electromagnetic field provided by the Lagrangian coordinates will be a useful alternative technique for solving Maxwell's equations, in particular when the theory is extended to non-trivial media and sources. Certainly, approximation techniques based on sets of mixed

Eulerian and Lagrangian equations have proved valuable in the analogous non-relativistic spin 0 case (see references in [2]).

The hydrodynamic paths we have introduced do not form a Lorentz covariant structure. This is not a mathematical or conceptual defect – indeed, it is a property shared with electric and magnetic field lines, and energy flow lines derived from Poynting's vector. The difference here is that the latter sets of lines are derived from the Lorentz covariant electromagnetic theory whereas the new non-covariant hydrodynamic ones may be employed as a basis from which to derive the relativistic theory. The possibility of such derivation is not surprising. The absence of the fluid paths in the basic Eulerian equations suggests that $\rho(x)$ and $S(x)$ (and hence $\mathbf{E}(x)$ and $\mathbf{B}(x)$) may be regarded as "collective coordinates" – functions that describe the bulk properties of the system without depending on the complex details of the particulate substructure. Features peculiar to the Eulerian picture, such as Lorentz covariance, may therefore be viewed as collective rather than fundamental properties. If we regard the variables $q_0, \theta_0$ as playing a role analogous to those postulated in "hidden variables" interpretations in quantum mechanics, this conclusion may have implications for the hidden-variable programme where it has generally been felt that a Lorentz covariant model is a desirable goal. Clearly, if there is a version of the quantum theory to be interpreted that is not itself Lorentz covariant, there is less compulsion to impose that requirement on the hidden variables.